\documentclass{article}
\usepackage{colortbl}

\usepackage{PRIMEarxiv}
\usepackage{float}
\usepackage[utf8]{inputenc} 
\usepackage[T1]{fontenc}    
\usepackage{hyperref}       
\usepackage{url}            
\usepackage{booktabs}       
\usepackage{amsfonts}       
\usepackage{nicefrac}       
\usepackage{microtype}      
\usepackage{lipsum}
\usepackage{fancyhdr}       
\usepackage{graphicx}       
\graphicspath{{media/}}     
\usepackage[table]{xcolor}
\usepackage{array}

\definecolor{categorycolor}{HTML}{DFF0D8} 
\definecolor{labelcolor}{HTML}{D9EDF7} 
\title{6KSFx Synth Dataset
\thanks{\textit{\underline{Citation}}: 
\textbf{Garcia and Reiss. 6KSFx. orcID: 0009-0007-3986-4518}} 
}

\author{
  Nelly Garcia Sihuay\\
  Centre for Digital Music  \\
  Queen Mary University of London \\
  London\\
  \texttt{\ n.v.a.garcia-sihuay@qmul.ac.uk} \\
   \And
  Joshua Reiss \\
  Centre for Digital Music \\
  Queen Mary University of London \\
  London\\
  \texttt{joshua.reiss@qmul.ac.uk} \\
}

\begin{document}
\maketitle

\begin{abstract}
Procedural audio, often referred to as "digital Foley", generates sound from scratch using computational processes. It represents an innovative approach to sound-effects creation. However, the development and adoption of procedural audio has been constrained by a lack of publicly available datasets and models, which hinders evaluation and optimization. To address this important gap, this paper presents a dataset of 6000 synthetic audio samples specifically designed to advance research and development in sound synthesis within 30 sound categories. By offering a description of the diverse synthesis methods used in each sound category and supporting the creation of robust evaluation frameworks, this dataset not only highlights the potential of procedural audio, but also provides a resource for researchers, audio developers, and sound designers. This contribution can accelerate the progress of procedural audio, opening up new possibilities in digital sound design.
\end{abstract}

\keywords{synthesis methods \and dataset\and subtractive \and physically informed \and modal \and additive }

\section{Introduction}
Sound effects are defined as non-musical and non-dialogue audio elements. They are an integral part of audiovisual and audio-only projects \cite{foleys} and play an important role in tasks such as classification, generation and retrieval. However, the scarcity of publicly available datasets, coupled with the cost and time required to create custom sound libraries, is a significant barrier to reproducibility and innovation in the field.

Existing datasets, such as AudioSet \cite{AudioSet}, VGGSound \cite{vggsound}, UrbanSound8k \cite{urban}, and Clotho \cite{clotho}, consist of pre-recorded sounds with minimal or no synthetic samples. This limitation hinders the evaluation and optimization of sound synthesis methods. Moreover, auditory research has predominantly focused on speech and music, leaving sound effects comparatively underexplored.

Procedural audio is defined by the use of algorithms to generate sound dynamically in real time. It is a promising approach for creating synthetic samples. However, its adoption is constrained by concerns regarding its believability compared to pre-recorded sounds and the absence of standardized evaluation methods.
\\

This dataset seeks to address these challenges by:

\begin{itemize}
    \item Provide a public dataset containing both synthetic and pre-recorded samples.
    \item Document the synthesis methods used for each category, allowing for further optimization.
    \item Support the development of standardized evaluation methods for synthetic sound models.
\end{itemize}

While generative models are emerging as a promising option for creating sound effects, they often lack the necessary parameters for meaningful user interaction, limiting their usefulness in sound design. However, progress has been made in optimizing generative models for music, as demonstrated by works such as \cite{vampnet} and \cite{spec}. There is also some research into controllable audio generation, such as \cite{sketch}, \cite{sketchsynth} and \cite{Yipin}. Despite these advances, there remains a significant opportunity to develop the field of procedural audio with a dataset tailored to bridge the gap between generative capabilities and user control, allowing for more intuitive and creative sound design.

\section{Previous work}
\label{sec:headings}

Computational sound synthesis techniques, such as procedural audio, offer a dynamic and flexible approach to creating these effects. Unlike traditional audio methods that rely on pre recorded sound libraries, procedural audio generates sounds algorithmically in real-time. This paradigm shift in sound design aligns well with the growing demand for immersive and interactive experiences in fields like gaming, virtual reality, and film.

The term "sound effects" refers to distinctive sounds with easily identifiable characteristics, as described by \cite{Dafx}.

\textbf{Advantages of Procedural Audio}
\\
Procedural audio offers several key advantages over conventional approaches, making it an attractive choice for both researchers and practitioners. In the list below we mentioned some of the advantages of this method: 

\begin{itemize}
    \item \textbf{Diversity and Non-Repetition:} According to \cite{proc}, procedural audio removes the need for extensive pre-recorded sample libraries by enabling the creation of unique sound instances programming. This approach not only saves time but also enhances realism by replicating the natural variability found in acoustic environments, such as the subtle differences between individual footsteps or raindrops.
    \item \textbf{Reduced Memory Usage:} Generating sounds in real-time eliminates the need to store large audio files, significantly optimizing memory usage in resource-constrained systems like mobile devices and game consoles. As highlighted in \cite{Dmitris}, this advantage is particularly valuable in modern applications requiring lightweight yet high-quality audio solutions.
    \item \textbf{Interactivity:} Procedural audio dynamically responds to user input and real-time events, providing an interactive and immersive auditory experience. For example, sounds can seamlessly adapt to a player’s actions in a video game, as described in \cite{comp}, enhancing the overall sense of engagement.
\end{itemize}

\textbf{Challenges in Procedural Audio Research}

Despite its potential, procedural audio research faces notable challenges. A critical issue is the lack of a standardized evaluation framework, which limits the ability to compare techniques and assess their effectiveness. Previous studies, such as \cite{SFXSynth}, have attempted to evaluate perceptual qualities of sound textures like fire and rain across different synthesis methods. However, these efforts were constrained by their narrow scope and the unavailability of public datasets, hindering reproducibility and broader exploration.

Subjective evaluation methods, such as those discussed in \cite{Objective}, are occasionally employed in procedural audio studies. However, without datasets containing detailed synthesis metadata, these methods often lack the rigor and transparency necessary for consistent analysis. For instance, Farnell's behavioral abstraction framework \cite{farnell} distinguishes between physical models and physically informed models, yet validation datasets to test these distinctions are scarce. Similarly, works like \cite{percept} explore additive synthesis but faced limitations when synthetic samples were no longer publicly accessible.

\textbf{Contribution of This Dataset}

To address these limitations, the dataset presented in this study provides a comprehensive collection of synthetic sound samples categorized by synthesis method. This resource enables detailed comparisons and evaluations of procedural audio techniques, fostering reproducibility, and advancing research.

The synthesis methods represented in this dataset include the following:

\begin{enumerate}
    \item \textbf{Additive:} Constructs complex sounds by summing sine waves with unique frequency, amplitude, and phase parameters. Based on Fourier theory, this method is a foundational approach to generate harmonic tones \cite{Add}.
    \item \textbf{Subtractive:} Starts with harmonically rich waveforms (e.g., sawtooth or square) and removes specific frequency components using filters like low-pass and high-pass filters. Subtractive synthesis is widely used in music production and sound design \cite{sub}.
    \item \textbf{Granular:} Combines thousands of small audio fragments or "grains," each manipulable in pitch, duration, or frequency. This technique produces complex textures and evolving sounds.
    \item \textbf{Physical Modeling:} Simulates the physical properties of real-world instruments (e.g., strings, membranes) using mathematical models. It captures interactions between exciters (e.g., bows) and resonators (e.g., strings), providing realistic simulations of acoustic behaviors \cite{phys}.
    \item \textbf{Physically Informed:} Combines elements of physical modeling with digital signal processing to emulate acoustic behaviors while allowing creative non-physical modifications. This hybrid approach enhances flexibility and expressive potential \cite{phys2}.
    \item \textbf{Modal:} A specialized variant of physical modeling that represents vibrations as discrete resonant modes. This method excels at simulating resonances in materials and instruments with high precision \cite{phys2}.
    \item \textbf{Signal Modeling:} Focuses on analyzing and reconstructing sounds by extracting key parameters, such as pitch, envelope, and timbre. Techniques such as linear predictive coding (LPC) are common in this category \cite{synth2}.
    \item \textbf{Frequency Modeling:} Manipulates frequency-domain representations of audio signals. Frequency modulation (FM) synthesis and spectral modeling fall under this category and are widely used to produce rich tonal textures \cite{fm}.
\end{enumerate}

Sound synthesis methods can be combined to create a single sound sample, allowing greater flexibility and complexity in audio design. In this study, we analyzed the underlying code for each sound effect to identify the specific synthesis methods used. Using the categorization framework proposed by \cite{Dmitris}, we classified the sound categories based on the synthesis techniques used, as detailed in Section 3.

The creation and publication of synthetic sound samples addresses a critical need by providing a publicly accessible resource that includes diverse sound categories and detailed synthesis metadata. This dataset facilitates the development of standardized evaluation frameworks, improves reproducibility, and lays the groundwork for further research and innovation in procedural audio.

\section{Dataset}

This dataset was constructed using 12,000 five-second sound samples, evenly distributed across 30 categories of real and synthetic audio. The sound categories were carefully chosen for their accessibility and relevance to common use cases in audiovisual projects such as film, games, virtual reality and interactive media. These categories cover a wide range of sounds, from environmental noises to mechanical and synthetic effects, ensuring the applicability of the dataset to a wide range of applications.

Real samples were sourced from online sound libraries, known for their extensive collections and high-quality recordings including the BBC \footnote{https://sound-effects.bbcrewind.co.uk}, Hybrid \footnote{https://www.prosoundeffects.com}, Soundsnap \footnote{https://www.soundsnap.com}, and Pixabay \footnote{https://pixabay.com}. These were paired with synthetic samples generated using the Nemisindo \footnote{https://nemisindo.com} online procedural audio engine. Due to copyright restrictions, the real samples are not publicly available. However, links to the providers and their corresponding sound categories are provided. On the other hand, \textbf{We are releasing 6,000 synthetic samples as a public dataset}, a total of 2.56 GB and divided into 30 samples per category available on Zenodo\footnote{\url{https://zenodo.org/records/14517916}}. Each sample is named in the following format:  
\begin{center}
\textit{"Name of the sound category" - "Sample number" - "Label"}
\end{center}

\begin{figure}[H] \centering \includegraphics[width=0.8\linewidth]{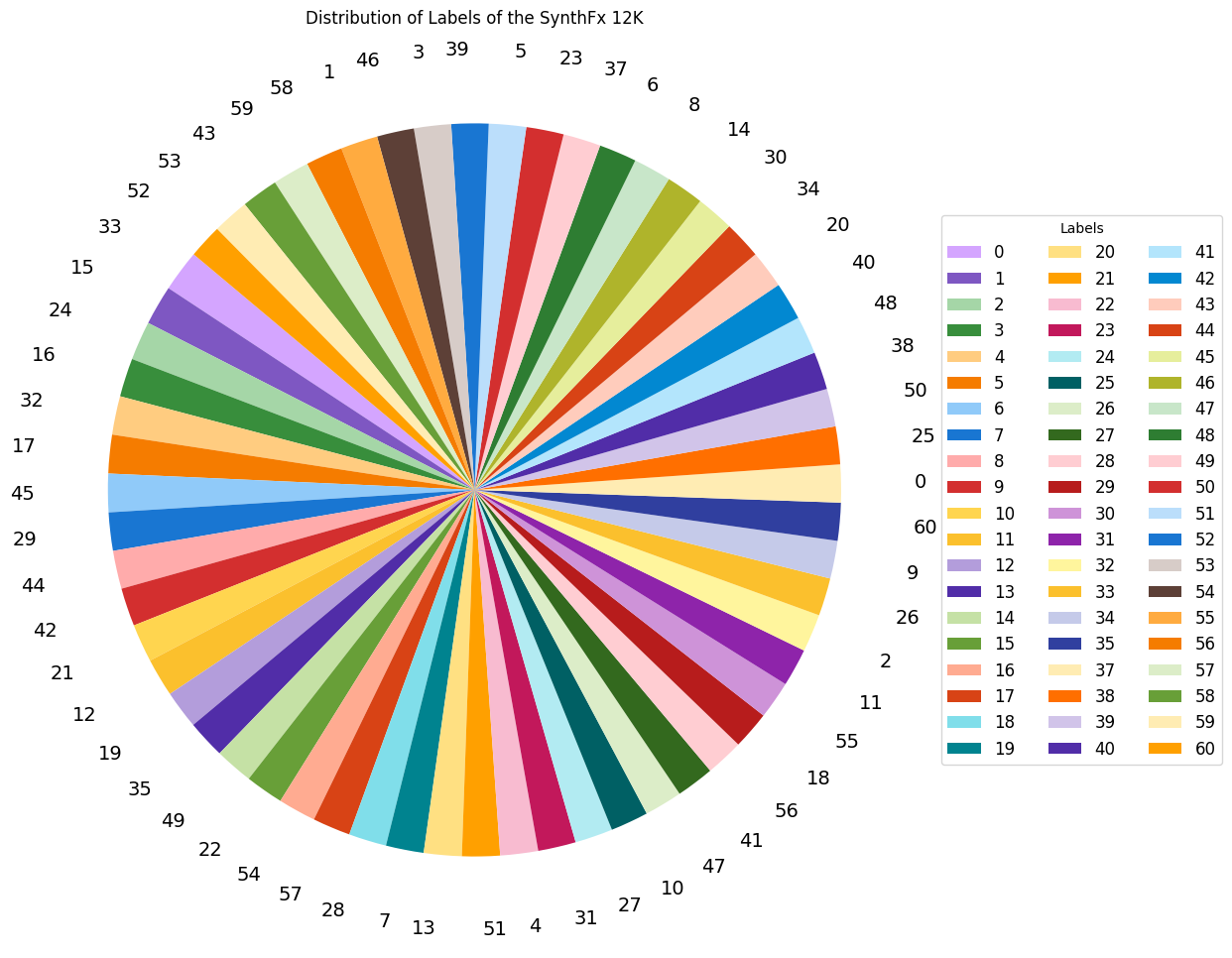} \caption{6KSFx Dataset Distribution: Soft colors = recorded sounds. Dark colors = synthetic sounds.} \label{dist} \end{figure}

To ensure a balanced representation across categories, each category accounts for 1.8\% of the dataset (200 samples per category), as shown in Figure \ref{dist}.

We prioritize the balance to enable fair comparisons between real and synthetic audio in evaluation tasks and training machine learning models. The distribution shown in \ref{dist} represents the 30 sound categories, with real samples represented by lighter tones and synthetic samples by darker tones (e.g., 0 is the real sample category of applause while 1 is the synthetic sample category of applause), the label distribution is described in Table \ref{labels}, highlighted in blue are the labels for the synthetic samples, and with green the sound category name.

\begin{table}[h]
    \centering
    \begin{tabular}{|>{\columncolor{categorycolor}}c|c|>{\columncolor{labelcolor}}c|>{\columncolor{categorycolor}}c|c|>{\columncolor{labelcolor}}c|}
        \hline
        \textbf{Sound Category} & \textbf{R.S. Label} & \textbf{S.S. Label} & \textbf{Sound Category} & \textbf{R.S Label} & \textbf{S.S. Label} \\
        \hline
        \textit{Applause} & 0 & 1 & \textit{Wind} & 30 & 31 \\
        \hline
        \textit{Church Bells} & 2 & 3 & \textit{Boat} & 32 & 33 \\
        \hline
        \textit{Bubbles} & 4 & 5 & \textit{Jet} & 34 & 35 \\
        \hline
        \textit{Droplets} & 6 & 7 & \textit{Whoosh} & 37 & 38 \\
        \hline
        \textit{Crackling} & 8 & 9 & \textit{Fireworks} & 39 & 40 \\
        \hline
        \textit{Glass Debris} & 10 & 11 & \textit{Concrete Footsteps} & 41 & 42 \\
        \hline
        \textit{Engine} & 12 & 13 & \textit{Wood Footsteps} & 42 & 43 \\
        \hline
        \textit{Explosions} & 14 & 15 & \textit{Space Cannon} & 44 & 45 \\
        \hline
        \textit{Fire Embers} & 16 & 17 & \textit{Spray} & 46 & 47 \\
        \hline
        \textit{Gunshot} & 18 & 19 & \textit{Metal Debris} & 48 & 49 \\
        \hline
        \textit{Helicopter} & 20 & 21 & \textit{Rocket} & 50 & 51 \\
        \hline
        \textit{Pouring Hot Water} & 22 & 23 & \textit{Concrete Debris} & 52 & 53 \\
        \hline
        \textit{Rain} & 24 & 25 & \textit{Twang} & 54 & 55 \\
        \hline
        \textit{Thunder} & 26 & 27 & \textit{Bounce (Rubber)} & 56 & 57 \\
        \hline
        \textit{Waves} & 28 & 29 & \textit{Electric Buzz} & 58 & 59 \\
        \hline
    \end{tabular}
    \caption{Labels per sound category. R.S = Real sample. S.S = Synthetic sample.}
    \label{labels}
\end{table}

For added realism and making the comparison with real samples fair, the synthetic samples were processed using reverb and equalization adjustments. Each sample is 5 seconds long, recorded at 44.1 kHz,  mono, and 16 bits per sample. Pairs of samples (e.g., 0 and 1, 2 and 3) represent the same sound category. The complete preprocess code, as well as the label distribution information, is available on Github\footnote{\url{https://github.com/nellyngz95/6KSFX.git}}.

Following the narrative of Section 2, we analyze the code of each sound model. Table \ref{synthmethods} provides an overview of the sound synthesis methods used for each category and their corresponding labels. Please note that classifications can be subjective. This broader taxonomy has certain simplifications, but generally covers the methods used in each synthesis model.

\begin{table}[H]
\centering
\renewcommand{\arraystretch}{1.5} 
\setlength{\tabcolsep}{6pt} 
\begin{tabular}{|>{\columncolor{categorycolor}}c|c|c|c|>{\columncolor{labelcolor}}c|}
\hline
\textbf{Sound Category} & \textbf{Method 1} & \textbf{Method 2} & \textbf{Method 3} & \textbf{Label} \\ \hline
\textit{Applause  }              & Additive          & -               & -               & 1             \\ \hline
\textit{Church Bells}            & Additive          & Phys. Informed    & Phys. Modeling    & 3             \\ \hline
\textit{Bubbles   }              & Additive          & Phys. Modeling    & Modal             & 5             \\ \hline
\textit{Droplets }               & Additive          & Phys. Informed    & Modal             & 7             \\ \hline
\textit{Crackling }              & P. Informed            & Granular            & -          & 9             \\ \hline
\textit{Debris Glass}            & Modal  & Granular               & -             & 11            \\ \hline
\textit{Engine}               & Add/Sinusoidal    & Phys. Modeling      & -            & 13            \\ \hline
\textit{Explosions}              & Subtractive       & Phys. Informed    & Modal             & 15            \\ \hline
\textit{Fire }                   & Additive          & Modal             & Phys. Informed    & 17            \\ \hline
\textit{Gunshot}                 & Phys. Informed    & Additive   & Phys. Modeling    & 19            \\ \hline
\textit{Helicopter}              & Phys. Informed               &  Phys. Modeling       & -               & 21            \\ \hline
\textit{Pouring Hot Water}       & Additive          & Phys. Modeling    & Subtractive       & 23            \\ \hline
\textit{Rain }                   & Additive          & Phys. Informed    & Modal             & 25            \\ \hline
\textit{Thunder}                 & Subtractive               & Phys. Informed              & -              & 27            \\ \hline
\textit{Waves}                   & Subtractive       & Add/Sinusoidal    & -              & 29            \\ \hline
\textit{Wind }                   & Phys. Informed              & Subtractive             & -            & 31            \\ \hline
\textit{Boat Engine}             & Additive          & -              & -               & 33            \\ \hline
\textit{Jet}                     & Additive          & Phys. Informed    & Modal             & 35            \\ \hline
\textit{Whoosh}                  & Frequency Mod.    & -            & -               & 37            \\ \hline
\textit{Fireworks}               & Additive               & Phys. Informed              & -            & 39            \\ \hline
\textit{Concrete Footsteps}      & Phys. Informed    & Add/Sinusoidal    & -               & 41            \\ \hline
\textit{Wood Footsteps}          & Phys. Informed    & Add/Sinusoidal    & -            & 43            \\ \hline
\textit{Space Cannon}            & Frequency Mod.    & Modal             & Subtractive       & 45            \\ \hline
\textit{Spray}                   & Subtractive               & Phys. Informed             & -             & 47            \\ \hline
\textit{Metal Debris}        & Signal Modeling   & Granular               & -                  & 49            \\ \hline
\textit{Rocket }                 & Modal             & Subtractive       & Signal Based      & 51            \\ \hline
\textit{Concrete Debris}       & Signal Modeling   & Granular               & -               & 53            \\ \hline
\textit{Twang}                   & Phys. Modeling              & Phys. Informed               & -             & 55            \\ \hline
\textit{Bounce Rubber}           & Phys. Informed    & Add/Sinusoidal    & -             & 57            \\ \hline
\textit{Electric Buzz}           & Subtractive       & -              & -            & 59            \\ \hline
\end{tabular}
\caption{Types of sound synthesis methods used across sound categories. 
'-' = Not Applicable, Phys. = Physically}
\label{synthmethods}
\end{table}

\section{Conclusion}
Sound synthesis represents an opportunity for innovation in the creation of sound samples, offering unprecedented flexibility and control in audio design. However, the lack of accessible, high-quality synthetic datasets has slowed down the progress in research, the development of standardized evaluation frameworks, and the wider adoption of procedural audio techniques. To address this challenge, we have launched a dataset of 6,000 carefully crafted synthetic samples, categorized by synthesis method. 

By integrating both real and synthetic sounds, our dataset creates a unique bridge between traditional recordings and advanced procedural audio methods. This duality supports multiple use cases, from training and testing machine learning algorithms to exploring comparative evaluations of audio quality. 

Beyond its immediate applications, the dataset aims to catalyze wider advances in sound synthesis by encouraging the development of standardized evaluation methodologies.These frameworks will enable researchers to better assess the realism, diversity and quality of synthetic audio, while fostering collaboration across disciplines.

It will serve as a tool for researchers, audio developers, and sound designers to evaluate, benchmark, and refine sound synthesis techniques. This dataset can be used to expand current sound effect datasets. Ultimately, this dataset aims to drive progress in procedural audio research, inspire the creation of new tools and techniques, and advance the state of the art in sound synthesis for the benefit of academia, industry, and the creative community alike.

\bibliographystyle{plain} 

\bibliography{references} 

\end{document}